\documentclass[10pt]{article}
\usepackage{graphicx}
\usepackage{amsmath}
\usepackage{amssymb}
\usepackage{caption2}
\begin{document}
\begin{center}
{\large {\bf \sc{  Analysis of the triply-charm tetraquark states via the QCD sum rules }}} \\[2mm]
Zhi-Gang  Wang \footnote{E-mail: zgwang@aliyun.com.  }     \\
 Department of Physics, North China Electric Power University, Baoding 071003, P. R. China
\end{center}

\begin{abstract}
In this work,  we construct the diquark-antidiquark type local currents to interpolate the triply-charm tetraquark states, and investigate
  the spectroscopy of the scalar, axialvector and tensor triply-charm tetraquark states via the QCD sum rules comprehensively. As a  byproduct, we  obtain the masses  of four vector triply-charm tetraquark states.
 \end{abstract}



\section{Introduction}
In 2003, the  Belle collaboration   observed   a narrow charmonium-like structure  $X(3872)$ in the $\pi^+ \pi^- J/\psi$ invariant mass distribution in the decays  $B^\pm \to K^\pm \pi^+ \pi^- J/\psi$  \cite{X3872-2003}, which cannot be identified  as the $\chi_{c1}(\rm 2P)$ state in the traditional quark model. Thereafter, a number of exotic states  were observed, such as the $Z_c^{\pm}(3900)$ in the $J/\psi\pi^\pm$ invariant mass distributions by the BESIII and Belle collaborations \cite{BES3900,Belle3900}, the $P_c(4312)$, $P_c(4380)$, $P_c(4440)$ and $P_c(4457)$ ($P_{cs}(4338)$ and $P_{cs}(4459)$) in the $J/\psi p$ ($J/\psi\Lambda$) invariant mass distribution by the LHCb collaboration \cite{LHCb-4380,LHCb-Pc4312}(\cite{LHCb-Pcs4338,LHCb-Pcs4459-2021}),
the $X(4140)$, $X(4274)$, $X(4500)$ and $X(4700)$ in the $J/\psi \phi$ invariant  mass distribution by the LHCb collaboration \cite{LHCb-16061,LHCb-16062},
the $T_{c\bar{s}}(2900)$ in the $D_s^+ \pi^+$ and $ D_s^+ \pi^-$ invariant mass distributions by the LHCb collaboration \cite{LHCbTcs2900-1,LHCbTcs2900-2},
the $X(6660)$, $X(6900)$ and $X(7100)$ in the $J/\psi J/\psi$ invariant mass distribution by the ATLAS, CMS and LHCb collaborations \cite{LHCb-cccc-2006,ATLAS-cccc-2023,CMS-cccc-2023}, the $T_{cc}^+(3875)$ in the $D^0D^0\pi^+$  invariant mass distribution by the LHCb collaboration \cite{LHCb-Tcc,LHCb-Tcc-detail},
which cannot be accommodated in the traditional quark models.
There have seen several possible interpretations, such as the tetraquark/pentaquark states, hadronic molecular states, dynamically generated resonances,  kinematical effects, etc \cite{HXChen-review-1601, RFLebed-review-1610,AEsposito-review-1611,FKGuo-review-1705, AAli-review-1706,SLOlsen-review-1708,MNielsen-review-1812,
YRLiu-review-1903,CPShen-review-1907,LiuXH-review-2020,WangBo-review-2023,
LiuMZ-review-2025,WangZG-review-2026}. In the scenarios of multiquark states and molecular states, there exist four or five valence quarks, which are the most popular interpretations.

In 2013, the LHCb collaboration observed the decays $B^+_c \rightarrow J/\psi D_s^+$ and $ J/\psi D_s^{*+}$ with the statistical significance  in excess of 9 standard deviations for the first time \cite{LHCb-JpsiD-PRD-2013}. In 2016 (also in 2022), the ATLAS collaboration measured several ratios among the branching fractions involving the  decays $B_c^+ \to J/\psi D_s^+$ and $J/\psi D_s^{*+}$ \cite{ATLAS-JpsiD-EPJC-2016} (\cite{ATLAS-JpsiD-JHEP-2022}).
In 2017, the LHCb collaboration observed  the decays $B_c^+ \to J/\psi D^0 K^+$ and $J/\psi D^{*0} K^+$ for the first time, and reported the first evidence  for the decays $B_c^+ \to J/\psi D^{*+} K^{*0}$ and $ J/\psi D^+ K^{*0}$  \cite{LHCb-JpsiD-PRD-2017}.
Very recently, the LHCb collaboration observed  the decays $B^+_c \to J/\psi D^{\ast+}$ with the significance of 5.4 standard deviations for the first time, and measured the relative branching fraction \cite{LHCb-JpsiDv-2026},
\begin{eqnarray}
\frac{\mathcal{B}(B^+_c \to J/\psi D^{\ast+})}{\mathcal{B}(B^+_c \to J/\psi D^+_s)} &=& 0.067 \pm 0.016 \pm 0.003 \, .
\end{eqnarray}
Analogously, we expect that some exotic structures in the $J/\psi D$, $J/\psi D^*$, $J/\psi D_s$ and $J/\psi D_s^*$ invariant mass distributions  might be observed in the future, which would have the valence quarks $cc\bar{c}\bar{q}$ with $q=u$, $d$ or $s$.

Theoretically, there have been a number of works on the triply-heavy tetraquark states based on the potential quark models \cite{Tetra-ccc-QuarkModel-ChenK-EPJA-2017,Tetra-ccc-QuarkModel-LuQF-PRD-2021,
 Tetra-ccc-QuarkModel-HuangHX-PRD-2023,Tetra-ccc-QuarkModel-DengWZ-PRD-2023,
 Tetra-ccc-QuarkModel-Mutuk-EPJC-2023,Tetra-ccc-QuarkModel-YangG-PRD-2024,
 Tetra-ccc-QuarkModel-JiaDJ-EPJC-2024,Tetra-ccc-QuarkModel-Galkin-PPN-2025,
 Tetra-ccc-QuarkModel-AnHT-EPJC-2026} and  QCD sum rules \cite{Tetra-ccc-QCDSR-ChenW-PRD-2017,Tetra-ccc-QCDSR-ZhangWS-NPA-2025,
 Tetra-ccc-QCDSR-LiuYL-2506,Tetra-ccc-QCDSR-ZhangWS-2608}, but the predictions differ from each other significantly. In the present work, we would like to focus on the QCD sum rules.

The QCD sum rules approach is a robust tool to investigate  the exotic states, and has been applied extensively to calculate the masses, pole residues, hadronic coupling constants, etc \cite{MNielsen-review-1812,WangZG-review-2026}. We usually choose the color antitriplet-triplet type, sextet-antisextet type, singlet-singlet type and octet-octet  type local four-quark currents to investigate the tetraquark states, and we usually call the singlet-singlet type tetraquark states as the hadronic molecular states according to the two color-neutral clusters without requiring exchanging intermediate bosons.

To my knowledge,  most works are devoted to the diquark-antidiquark type tetraquark states in the QCD sum rules \cite{WangZG-review-2026}, where we take the diquark operators in color antitriplet as the fundamental   constituents to construct  the local currents.  The one-gluon exchange induces attractive (repulsive) interaction in  color  antitriplet (sextet), which favors (disfavors)  formation of two-quark correlations  in color antitriplet (sextet) \cite{One-gluon-1,One-gluon-2}, and we would like to adopt the diquark operators in  color antitriplet as the fundamental  constituents as usual to construct the four-quark currents.

The diquark operators $\varepsilon^{abc}q^{T}_b C\Gamma q^{\prime}_c$  have  five  structures, where $C\Gamma=C\gamma_5$, $C$, $C\gamma_\mu \gamma_5$,  $C\gamma_\mu $ and $C\sigma_{\mu\nu}$ (or $C\sigma_{\mu\nu}\gamma_5$) for the scalar ($S$), pseudoscalar ($P$), vector ($V$), axialvector ($A$)  and  tensor ($T$) diquarks, respectively, the $a$, $b$, $c$ are color indexes. The tensor diquarks have both spin-parity $J^P=1^+$ and $1^-$ components, we  represent  them as $\widetilde{A}$ and $\widetilde{V}$, respectively. The negative parity implies there exists an additional P-wave, therefore the two-quark correlation would be weak and unstable.   In this work, we would like to investigate  the ground state mass spectrum of the diquark-antidiquark type triply-charm tetraquark states using the stable diquarks  via the QCD sum rules comprehensively.

The article is arranged as:  we obtain the QCD sum rules for  the triply-charm tetraquark states in section 2; in section 3, we   present the numerical results and discussions; section 4 is reserved for our conclusion.

\section{QCD sum rules for  the  triply-charm  tetraquark states}

As the first step, we write down  the  correlation functions $\Pi(p)$, $\Pi_{\mu\nu}(p)$ and $\Pi_{\mu\nu\alpha\beta}(p)$,
\begin{eqnarray}\label{CF-Pi}
\Pi(p)&=&i\int d^4x e^{ip \cdot x} \langle0|T\Big\{J(x)J^{\dagger}(0)\Big\}|0\rangle \, ,\nonumber\\
\Pi_{\mu\nu}(p)&=&i\int d^4x e^{ip \cdot x} \langle0|T\Big\{J_\mu(x)J_{\nu}^{\dagger}(0)\Big\}|0\rangle \, ,\nonumber\\
\Pi_{\mu\nu\alpha\beta}(p)&=&i\int d^4x e^{ip \cdot x} \langle0|T\Big\{J_{\mu\nu}(x)J_{\alpha\beta}^{\dagger}(0)\Big\}|0\rangle \, ,
\end{eqnarray}
where $J(x)=J_{AA}(x)$, $J_{\widetilde{A}\widetilde{A}}(x)$, $J_\mu(x)=J^{AA}_{\mu}(x)$, $J_{\mu}^{\widetilde{A}\tilde{A}}(x)$,
$J_{\mu\nu}(x)=J^{AA}_{1,\mu\nu}(x)$, $J^{\tilde{A}\tilde{A}}_{1,\mu\nu}(x)$, $J^{TS}_{1,\mu\nu}(x)$, $J^{AA}_{2,\mu\nu}(x)$, $J^{\tilde{A}\tilde{A}}_{2,\mu\nu}(x)$,
\begin{eqnarray}
J_{AA}(x)&=&\varepsilon^{ijk}\varepsilon^{imn}c^{T}_j(x)C\gamma_\mu c_k(x)\,  \bar{c}_m(x)\gamma^\mu C \bar{q}^{T}_n(x) \, ,\nonumber \\
J_{\tilde{A}\tilde{A}}(x)&=&\varepsilon^{ijk}\varepsilon^{imn}c^{T}_j(x)
C\sigma_{\mu\theta}  c_k(x) \, \bar{c}_m(x)\sigma^{\mu\theta} C \bar{q}^{T}_n(x) \, ,
\end{eqnarray}
\begin{eqnarray}
J^{AS}_\mu(x)&=&\varepsilon^{ijk}\varepsilon^{imn}c^{T}_j(x)C\gamma_\mu c_k(x)\,  \bar{c}_m(x)\gamma_5 C \bar{q}^{T}_n(x) \, ,\nonumber \\
J^{\tilde{A}A}_\mu(x)&=&\varepsilon^{ijk}\varepsilon^{imn}c^{T}_j(x)C\sigma_{\mu\theta}\gamma_5 c_k(x) \, \bar{c}_m(x)\gamma^\theta C \bar{q}^{T}_n(x) \, ,
\end{eqnarray}
\begin{eqnarray}
J^{AA}_{1,\mu\nu}(x)&=&\frac{\varepsilon^{ijk}\varepsilon^{imn}}{\sqrt{2}}\left[c^{T}_j(x)C\gamma_\mu c_k(x)\,  \bar{c}_m(x)\gamma_\nu C \bar{q}^{T}_n(x)-c^{T}_j(x)C\gamma_\nu c_k(x)\,  \bar{c}_m(x)\gamma_\mu C \bar{q}^{T}_n(x) \right] \, ,\nonumber \\
J^{\tilde{A}\tilde{A}}_{1,\mu\nu}(x)&=&\frac{\varepsilon^{ijk}\varepsilon^{imn}}{\sqrt{2}}
\left[c^{T}_j(x)C\sigma_{\mu}{}^\theta c_k(x)\,  \bar{c}_m(x)\sigma_{\nu\theta} C \bar{q}^{T}_n(x)-c^{T}_j(x)C\sigma_\nu{}^\theta c_k(x)\,  \bar{c}_m(x)\sigma_{\mu\theta} C \bar{q}^{T}_n(x) \right] \, ,\nonumber \\
J^{TS}_{1,\mu\nu}(x)&=&\varepsilon^{ijk}\varepsilon^{imn}
c^{T}_j(x)C\sigma_{\mu\nu} c_k(x)\,  \bar{c}_m(x)\gamma_{5} C \bar{q}^{T}_n(x) \, ,
\end{eqnarray}
\begin{eqnarray}
J^{AA}_{2,\mu\nu}(x)&=&\frac{\varepsilon^{ijk}\varepsilon^{imn}}{\sqrt{2}}\left[c^{T}_j(x)C\gamma_\mu c_k(x)\,  \bar{c}_m(x)\gamma_\nu C \bar{q}^{T}_n(x)+c^{T}_j(x)C\gamma_\nu c_k(x)\,  \bar{c}_m(x)\gamma_\mu C \bar{q}^{T}_n(x) \right] \, ,\nonumber \\
J^{\tilde{A}\tilde{A}}_{2,\mu\nu}(x)&=&\frac{\varepsilon^{ijk}\varepsilon^{imn}}{\sqrt{2}}
\left[c^{T}_j(x)C\sigma_{\mu}{}^\theta c_k(x)\,  \bar{c}_m(x)\sigma_{\nu\theta} C \bar{q}^{T}_n(x)+c^{T}_j(x)C\sigma_\nu{}^\theta c_k(x)\,  \bar{c}_m(x)\sigma_{\mu\theta} C \bar{q}^{T}_n(x) \right] \, ,\nonumber \\
\end{eqnarray}
 $q=u$, $d$, $s$, and we would like to take the isospin limit, the $i$, $j$, $k$, $m$, $n$ are  color indexes. For the doubly-charm diquarks, only the operators $\varepsilon^{ijk}c^T_jC\gamma_\mu c_k$, $\varepsilon^{ijk}c^T_jC\sigma_{\mu\nu} c_k$ and $\varepsilon^{ijk}c^T_jC\sigma_{\mu\nu} \gamma_5c_k$ could exist due to Fermi-Dirac statistics. For the singly-charm anti-diquarks, we choose the operators $\varepsilon^{ijk}\bar{c}_j\gamma_5 C \bar{q}^T_k$, $\varepsilon^{ijk}\bar{c}_j\gamma_\mu C \bar{q}^T_k$ and $\varepsilon^{ijk}\bar{c}_j\sigma_{\mu\nu} C \bar{q}^T_k$, as the operators
 $\varepsilon^{ijk}\bar{c}_j C \bar{q}^T_k$ and $\varepsilon^{ijk}\bar{c}_j\gamma_5\gamma_\mu C \bar{q}^T_k$ have negative parity and would couple potentially to the diquark states with larger masses.

Under parity transform $\widehat{P}$, the currents have the  properties,
\begin{eqnarray}
\widehat{P} J(x)\widehat{P}^{-1}&=&+J(\tilde{x}) \, , \nonumber\\
\widehat{P} J_\mu(x)\widehat{P}^{-1}&=&-J^\mu(\tilde{x}) \, , \nonumber\\
\widehat{P} J^{TS}_{\mu\nu}(x)\widehat{P}^{-1}&=&-J_{TS}^{\mu\nu}(\tilde{x}) \, , \nonumber\\
\widehat{P} J^{AA/\tilde{A}\tilde{A}}_{\mu\nu}(x)\widehat{P}^{-1}&=&+J_{AA/\tilde{A}\tilde{A}}^{\mu\nu}(\tilde{x}) \, ,
\end{eqnarray}
where  $x^\mu=(t,\vec{x})$ and $\tilde{x}^\mu=(t,-\vec{x})$.

\begin{table}
\begin{center}
\begin{tabular}{|c|c|c|c|c|c|c|c|c|}\hline\hline
 $Z_{cc\bar{c}}$              & $J^{P}$  & Currents       \\ \hline

$[cc]_{A}[\overline{cq}]_{A}$      & $0^{+}$  & $J_{AA}(x)$       \\

$[cc]_{\tilde{A}}[\overline{cq}]_{\tilde{A}}$                                     & $0^{+}$  & $J_{\widetilde{A}\widetilde{A}}(x)$        \\
 \hline

$[cc]_{A}[\overline{cq}]_{S}$      & $1^{+}$  & $J_{AS}^\mu(x)$       \\

$[cc]_{\tilde{A}}[\overline{cq}]_{A}$      & $1^{+}$  & $J_{\tilde{A}A}^\mu(x)$       \\

$[cc]_{A}[\overline{cq}]_{A}$              & $1^{+}$  & $J_{AA}^{1,\mu\nu}(x)$       \\

$[cc]_{\tilde{A}}[\overline{cq}]_{\tilde{A}}$              & $1^{+}$  & $J_{\tilde{A}\tilde{A}}^{1,\mu\nu}(x)$       \\

$[cc]_{\tilde{A}}[\overline{cq}]_{S}$                      & $1^{+}$  & $J_{TS}^{1,\mu\nu}(x)$       \\ \hline

$[cc]_{A}[\overline{cq}]_{A}$              & $1^{-}$  & $J_{AA}^{1,\mu\nu}(x)$       \\

$[cc]_{\tilde{A}}[\overline{cq}]_{\tilde{A}}$              & $1^{-}$  & $J_{\tilde{A}\tilde{A}}^{1,\mu\nu}(x)$       \\

$[cc]_{\tilde{V}}[\overline{cq}]_{S}$                      & $1^{-}$  & $J_{TS}^{1,\mu\nu}(x)$       \\ \hline

$[cc]_{A}[\overline{cq}]_{A}$              & $2^{+}$  & $J_{AA}^{2,\mu\nu}(x)$       \\

$[cc]_{\tilde{A}}[\overline{cq}]_{\tilde{A}}$              & $2^{+}$  & $J_{\tilde{A}\tilde{A}}^{2,\mu\nu}(x)$       \\

\hline\hline
\end{tabular}
\end{center}
\caption{ The quark structures and  currents  for the triply-charm tetraquark states. }\label{Current-Table}
\end{table}

On the hadron side, we  insert  a complete set of intermediate hadronic states with
the same quantum numbers as the four-quark currents $J(x)$, $J_\mu(x)$ and $J_{\mu\nu}(x)$ into the
correlation functions $\Pi(p)$, $\Pi_{\mu\nu}(p)$ and $\Pi_{\mu\nu\alpha\beta}(p)$   to obtain the hadronic representation
\cite{SVZ79-1,SVZ79-2,Reinders85}, and isolate the ground state triply-charm tetraquark states,
\begin{eqnarray}
\Pi(p)&=&\frac{\lambda_{Z^+}^2}{M_{Z^+}^2-p^2} +\cdots \nonumber\\
&=&\Pi_{+}(p^2) \, ,\nonumber\\
\Pi_{\mu\nu}(p)&=&\frac{\lambda_{Z^+}^2}{M_{Z^+}^2-p^2}\left( -g_{\mu\nu}+\frac{p_{\mu}p_{\nu}}{p^2}\right) +\cdots \nonumber\\
&=&\Pi_{+}(p^2)\left( -g_{\mu\nu}+\frac{p_{\mu}p_{\nu}}{p^2}\right)+\cdots \, ,\nonumber\\
\Pi^{AA/\tilde{A}\tilde{A},1}_{\mu\nu\alpha\beta}(p)&=&\frac{\lambda_{ Z^+}^2}{M_{Z^+}^2\left(M_{Z^+}^2-p^2\right)}\left(p^2g_{\mu\alpha}g_{\nu\beta} -p^2g_{\mu\beta}g_{\nu\alpha} -g_{\mu\alpha}p_{\nu}p_{\beta}-g_{\nu\beta}p_{\mu}p_{\alpha}+g_{\mu\beta}p_{\nu}p_{\alpha}+g_{\nu\alpha}p_{\mu}p_{\beta}\right) \nonumber\\
&&+\frac{\lambda_{ Z^-}^2}{M_{Z^-}^2\left(M_{Z^-}^2-p^2\right)}\left( -g_{\mu\alpha}p_{\nu}p_{\beta}-g_{\nu\beta}p_{\mu}p_{\alpha}+g_{\mu\beta}p_{\nu}p_{\alpha}+g_{\nu\alpha}p_{\mu}p_{\beta}\right) +\cdots  \nonumber\\
&=&\widetilde{\Pi}_{+}(p^2)\left(p^2g_{\mu\alpha}g_{\nu\beta} -p^2g_{\mu\beta}g_{\nu\alpha} -g_{\mu\alpha}p_{\nu}p_{\beta}-g_{\nu\beta}p_{\mu}p_{\alpha}+g_{\mu\beta}p_{\nu}p_{\alpha}+g_{\nu\alpha}p_{\mu}p_{\beta}\right) \nonumber\\
&&+\widetilde{\Pi}_{-}(p^2)\left( -g_{\mu\alpha}p_{\nu}p_{\beta}-g_{\nu\beta}p_{\mu}p_{\alpha}+g_{\mu\beta}p_{\nu}p_{\alpha}+g_{\nu\alpha}p_{\mu}p_{\beta}\right) \, ,\nonumber\\
\Pi^{TS,1}_{\mu\nu\alpha\beta}(p)&=&\frac{\lambda_{ Z^-}^2}{M_{Z^-}^2\left(M_{Z^-}^2-p^2\right)}\left(p^2g_{\mu\alpha}g_{\nu\beta} -p^2g_{\mu\beta}g_{\nu\alpha} -g_{\mu\alpha}p_{\nu}p_{\beta}-g_{\nu\beta}p_{\mu}p_{\alpha}+g_{\mu\beta}p_{\nu}p_{\alpha}+g_{\nu\alpha}p_{\mu}p_{\beta}\right) \nonumber\\
&&+\frac{\lambda_{ Z^+}^2}{M_{Z^+}^2\left(M_{Z^+}^2-p^2\right)}\left( -g_{\mu\alpha}p_{\nu}p_{\beta}-g_{\nu\beta}p_{\mu}p_{\alpha}+g_{\mu\beta}p_{\nu}p_{\alpha}+g_{\nu\alpha}p_{\mu}p_{\beta}\right) +\cdots  \nonumber\\
&=&\widetilde{\Pi}_{-}(p^2)\left(p^2g_{\mu\alpha}g_{\nu\beta} -p^2g_{\mu\beta}g_{\nu\alpha} -g_{\mu\alpha}p_{\nu}p_{\beta}-g_{\nu\beta}p_{\mu}p_{\alpha}+g_{\mu\beta}p_{\nu}p_{\alpha}+g_{\nu\alpha}p_{\mu}p_{\beta}\right) \nonumber\\
&&+\widetilde{\Pi}_{+}(p^2)\left( -g_{\mu\alpha}p_{\nu}p_{\beta}-g_{\nu\beta}p_{\mu}p_{\alpha}+g_{\mu\beta}p_{\nu}p_{\alpha}+g_{\nu\alpha}p_{\mu}p_{\beta}\right) \, , \nonumber\\
\Pi_{\mu\nu\alpha\beta}^{AA/\tilde{A}\tilde{A},2}(p)&=&\frac{\lambda_{ Z^+}^2}{M_{Z^+}^2-p^2}\left( \frac{\widetilde{g}_{\mu\alpha}\widetilde{g}_{\nu\beta}+\widetilde{g}_{\mu\beta}\widetilde{g}_{\nu\alpha}}{2}-\frac{\widetilde{g}_{\mu\nu}\widetilde{g}_{\alpha\beta}}{3}\right) +\cdots \, \, , \nonumber \\
&=&\Pi_{+}(p^2)\left( \frac{\widetilde{g}_{\mu\alpha}\widetilde{g}_{\nu\beta}+\widetilde{g}_{\mu\beta}\widetilde{g}_{\nu\alpha}}{2}-\frac{\widetilde{g}_{\mu\nu}\widetilde{g}_{\alpha\beta}}{3}\right) +\cdots\, ,
\end{eqnarray}
where $\widetilde{g}_{\mu\nu}=g_{\mu\nu}-\frac{p_{\mu}p_{\nu}}{p^2}$.  We add the superscripts $AA$, $\tilde{A}\tilde{A}$ and $TS$ to represent the diquark-antidiquark constituents, and add the superscripts $1$ and $2$ to distinguish the spins of the tetraquark states. As there also exist contributions from the tetraquark states with negative parity, we add the superscripts and subscripts $\pm$ to distinguish   the parity.  The  pole residues  are routinely defined by
\begin{eqnarray}
 \langle 0|J(0)|Z_{cc\bar{c}}^+(p)\rangle &=&\lambda_{Z^+}\, , \nonumber\\
 \langle 0|J_\mu(0)|Z_{cc\bar{c}}^+(p)\rangle &=&\lambda_{Z^+}\varepsilon_\mu\, , \nonumber\\
  \langle 0|J_{1,\mu\nu}^{TS}(0)|Z_{cc\bar{c}}^-(p)\rangle &=& \frac{\lambda_{Z^-}}{M_{Z^-}} \, \varepsilon_{\mu\nu\alpha\beta} \, \varepsilon^{\alpha}p^{\beta}\, , \nonumber\\
 \langle 0|J_{1,\mu\nu}^{TS}(0)|Z_{cc\bar{c}}^+(p)\rangle &=&\frac{\lambda_{Z^+}}{M_{Z^+}} \left(\varepsilon_{\mu}p_{\nu}-\varepsilon_{\nu}p_{\mu} \right)\, , \nonumber\\
  \langle 0|J_{1,\mu\nu}^{AA/\tilde{A}\tilde{A}}(0)|Z_{cc\bar{c}}^+(p)\rangle &=& \frac{\lambda_{Z^+}}{M_{Z^+}} \, \varepsilon_{\mu\nu\alpha\beta} \, \varepsilon^{\alpha}p^{\beta}\, , \nonumber\\
 \langle 0|J_{1,\mu\nu}^{AA/\tilde{A}\tilde{A}}(0)|Z_{cc\bar{c}}^-(p)\rangle &=&\frac{\lambda_{Z^-}}{M_{Z^-}} \left(\varepsilon_{\mu}p_{\nu}-\varepsilon_{\nu}p_{\mu} \right)\, , \nonumber\\
  \langle 0|J_{2,\mu\nu}^{AA/\tilde{A}\tilde{A}}(0)|Z_{cc\bar{c}}^+(p)\rangle &=& \lambda_{Z^+}\, \varepsilon_{\mu\nu} \, ,
\end{eqnarray}
where the  $\varepsilon_{\mu/\alpha}$ and $\varepsilon_{\mu\nu}$ are the polarization vectors \cite{WangZG-review-2026}.   In Table \ref{Current-Table}, we illustrate  the quark structures and  currents for the triply-charm tetraquark states.

It is not difficult to perform Fierz transformations to re-arrange the currents $J(x)$, $J_\mu(x)$ and $J_{\mu\nu}(x)$ into a series of color singlet-singlet type currents, for example,
\begin{eqnarray}
J_\mu^{AS}&=&-\frac{1}{2}\bar{q}c\,\bar{c}\gamma_\mu\gamma_5c
-\frac{i}{2}\bar{q}
\sigma_{\mu\nu}c\,\bar{c}\gamma^\nu\gamma_5c-\frac{i}{2}\bar{q}i\gamma_5c\,
\bar{c}\gamma_\mu c+\frac{i}{2}\bar{q}
\sigma_{\mu\nu}\gamma_5c\,\bar{c}\gamma^\nu c \nonumber\\
&&+\frac{1}{2}\bar{q}\gamma_\mu\gamma_5c\,\bar{c}c+\frac{i}{2}\bar{q}
\gamma^\nu\gamma_5c\,\bar{c}\sigma_{\mu\nu}c+\frac{i}{2}\bar{q}\gamma_\mu c\,
\bar{c} i\gamma_5c-\frac{i}{2}\bar{q}\gamma^\nu
c\,\bar{c}\sigma_{\mu\nu}\gamma_5 c\, ,
\end{eqnarray}

\begin{eqnarray}
J_\mu^{\tilde{A}A}&=&-\frac{3i}{2}\bar{q}c\,\bar{c}\gamma_\mu\gamma_5c
+\frac{1}{2}\bar{q}
\sigma_{\mu\nu}c\,\bar{c}\gamma^\nu\gamma_5c-\frac{3}{2}\bar{q}i\gamma_5c\,
\bar{c}\gamma_\mu c+\frac{1}{2}\bar{q}
\sigma_{\mu\nu}\gamma_5c\,\bar{c}\gamma^\nu c \nonumber\\
&&-\frac{3i}{2}\bar{q}\gamma_\mu\gamma_5c\,\bar{c}c+\frac{1}{2}\bar{q}
\gamma^\nu\gamma_5c\,\bar{c}\sigma_{\mu\nu}c-\frac{3}{2}\bar{q}\gamma_\mu c\,
\bar{c} i\gamma_5c+\frac{1}{2}\bar{q}\gamma^\nu
c\,\bar{c}\sigma_{\mu\nu}\gamma_5 c\, ,
\end{eqnarray}
the component $\bar{d}i\gamma_5c\,
\bar{c}\gamma_\mu c$ ($\bar{d}\gamma^\nu
c\,\bar{c}\sigma_{\mu\nu}\gamma_5 c$) couples potentially to the $J/\psi D$ ($J/\psi D^*$) pair. Thus we cannot exclude the possibility that there  exist  two-meson scattering state contributions, if they have the same quantum numbers as the  interpolating currents, as  the quantum field theory does not prohibit such couplings. We choose local four-quark currents, the  four  valence  quarks lie at the same point, and such four-quark systems  would  have the average
spatial  magnitudes  just like  the traditional  mesons \cite{WangZG-review-2026}.  The physical (traditional) mesons are spatial extended objects and have average  mass radii $\sqrt{\langle r^2\rangle}\approx 0.5\sim1.0\,\rm{fm}$,  the
average  mass radii of the meson pairs would be larger $1\,\rm{fm}$, therefore they cannot be interpolated by the local currents. In the local limit,  the  meson pairs  lose their existence and merge into tetraquark states, and their collective  contributions  can be safely absorbed into the pole residues \cite{WangZG-review-2026,WangHuangtao-2014-PRD,WangZG-Landau}.

On the QCD side,  we accomplish  the operator product expansion using the full $q$ and $c$ quark propagators,
\begin{eqnarray}
S_{ij}(x)&=& \frac{i\delta_{ij}\!\not\!{x}}{ 2\pi^2x^4}
-\frac{\delta_{ij}m_q}{4\pi^2x^2}-\frac{\delta_{ij}\langle
\bar{q}q\rangle}{12} +\frac{i\delta_{ij}\!\not\!{x}m_q
\langle\bar{q}q\rangle}{48}-\frac{\delta_{ij}x^2\langle \bar{q}g_s\sigma Gq\rangle}{192}+\frac{i\delta_{ij}x^2\!\not\!{x} m_q\langle \bar{q}g_s\sigma
 Gq\rangle }{1152}\nonumber\\
&& -\frac{ig_s G^{a}_{\alpha\beta}t^a_{ij}(\!\not\!{x}
\sigma^{\alpha\beta}+\sigma^{\alpha\beta} \!\not\!{x})}{32\pi^2x^2} -\frac{\delta_{ij}x^4\langle \bar{q}q \rangle\langle g_s^2 GG\rangle}{27648}-\frac{1}{8}\langle\bar{q}_j\sigma^{\mu\nu}q_i \rangle \sigma_{\mu\nu}  +\cdots \, ,
\end{eqnarray}
\begin{eqnarray}
C_{ij}(x)&=&\frac{i}{(2\pi)^4}\int d^4k e^{-ik \cdot x} \left\{
\frac{\delta_{ij}}{\!\not\!{k}-m_c}
-\frac{g_sG^n_{\alpha\beta}t^n_{ij}}{4}\frac{\sigma^{\alpha\beta}(\!\not\!{k}+m_c)+(\!\not\!{k}+m_c)
\sigma^{\alpha\beta}}{(k^2-m_c^2)^2}\right.\nonumber\\
&&\left. -\frac{g_s^2 (t^at^b)_{ij} G^a_{\alpha\beta}G^b_{\mu\nu}(f^{\alpha\beta\mu\nu}+f^{\alpha\mu\beta\nu}+f^{\alpha\mu\nu\beta}) }{4(k^2-m_c^2)^5}+\cdots\right\} \, ,\nonumber\\
f^{\alpha\beta\mu\nu}&=&(\!\not\!{k}+m_c)\gamma^\alpha(\!\not\!{k}+m_c)\gamma^\beta(\!\not\!{k}+m_c)\gamma^\mu(\!\not\!{k}+m_c)\gamma^\nu(\!\not\!{k}+m_c)\, ,
\end{eqnarray}
and  $t^n=\frac{\lambda^n}{2}$, the $\lambda^n$ is the Gell-Mann matrix
\cite{Reinders85,WangHuangtao-2014-PRD,Pascual-1984}.
We add  the terms $\langle\bar{q}_j\sigma_{\mu\nu}q_i \rangle$  from Fierz transformations   of
$\langle q_i \bar{q}_j\rangle$ before performing time-ordered product to absorb the gluons  emitted from other quark lines to  compute  the mixed condensates   $\langle\bar{q}g_s\sigma G q\rangle$ \cite{WangHuangtao-2014-PRD}.   If each heavy quark line emits a gluon and each light quark line provides  a quark-antiquark  pair, we obtain a quark-gluon  operator $g_sGg_sG g_sG \bar{q}q$   of dimension 10, which is  of the order $\mathcal{O}(\alpha_s^{3/2})$. We would like to truncate the operator product expansion up to the quark-gluon operators of the order $\mathcal{O}(\alpha_s^k)$ with $k\leq 1$  as usual in the QCD sum rule for the tetraquark and pentaquark (molecular) states  \cite{WangZG-review-2026,WangHuangtao-2014-PRD,Wang-tetra-formula,
WangZG-Huang-mole-EPJC-2014,WangZG-mole-EPJC-2014,
Wang1508-EPJC,WangXW-SCPMA-2022,WZG-penta-IJMPA,WangZG-XinQ-EPJC-2026}. We only need to  compute  the vacuum condensates $\langle\bar{q}q\rangle$, $\langle\frac{\alpha_{s}GG}{\pi}\rangle$, $\langle\bar{q}g_{s}\sigma Gq\rangle$,
$\langle\bar{q}q\rangle \langle\frac{\alpha_{s}GG}{\pi}\rangle$, then we  obtain  the QCD spectral densities through   dispersion relation as usual.

We match  the hadron side with the QCD  side of the components $\Pi_{\pm}(p^2)$ and $p^2\widetilde{\Pi}_{\pm}(p^2)$  below the continuum thresholds   $s_0$ and perform Borel transformations   respecting  to
 $P^2=-p^2$ to reach   the  QCD sum rules:
\begin{eqnarray}\label{QCDSR}
\lambda^2_{Z^\pm}\, \exp\left(-\frac{M^2_{Z^\pm}}{T^2}\right)= \int_{9m_c^2}^{s_0} ds\, \rho_{QCD}(s) \, \exp\left(-\frac{s}{T^2}\right) \, ,
\end{eqnarray}
the QCD spectral densities $\rho_{QCD}(s)$ are available  by contacting the corresponding author  via  E-mail.

We differentiate   Eq.\eqref{QCDSR}  respecting to $\tau=\frac{1}{T^2}$,  and obtain the QCD sum rules for
 the masses of the   triply-charm tetraquark states $Z_{cc\bar{c}}$,
 \begin{eqnarray}
 M^2_{Z^\pm}&=& -\frac{\int_{9m_c^2}^{s_0} ds\frac{d}{d \tau}\,\rho_{QCD}(s)\exp\left(-\tau s \right)}{\int_{9m_c^2}^{s_0} ds \, \rho_{QCD}(s)\exp\left(-\tau s\right)}\, .
\end{eqnarray}

\section{Numerical results and discussions}
At the beginning points, we choose  the usually used  values of the  vacuum condensates
$\langle\bar{q}q \rangle=-(0.24\pm 0.01\, \rm{GeV})^3$,  $\langle\bar{s}s \rangle=(0.8\pm0.1)\langle\bar{q}q \rangle$,
 $\langle\bar{q}g_s\sigma G q \rangle=m_0^2\langle \bar{q}q \rangle$, $\langle\bar{s}g_s\sigma G s \rangle=m_0^2\langle \bar{s}s \rangle$,
$m_0^2=(0.8 \pm 0.1)\,\rm{GeV}^2$, $\langle \frac{\alpha_s
GG}{\pi}\rangle=0.012\pm0.004\,\rm{GeV}^4$    at the energy scale  $\mu=1\, \rm{GeV}$
\cite{SVZ79-1,SVZ79-2,Reinders85,Colangelo-Review}, and  choose the $\overline{MS}$  quark  masses $m_{c}(m_c)=(1.275\pm0.025)\,\rm{GeV}$
 and $m_s(\mu=2\,\rm{GeV})=(0.095\pm0.005)\,\rm{GeV}$
 from the Particle Data Group \cite{PDG}.
In addition,  we set $m_u=m_d=m_q=0$,  and consider  their energy-scale dependence   \cite{Narison-mix},
 \begin{eqnarray}
 \langle\bar{q}q \rangle(\mu)&=&\langle\bar{q}q\rangle({\rm 1 GeV})\left[\frac{\alpha_{s}({\rm 1 GeV})}{\alpha_{s}(\mu)}\right]^{\frac{12}{33-2n_f}}\, , \nonumber\\
 \langle\bar{s}s \rangle(\mu)&=&\langle\bar{s}s \rangle({\rm 1 GeV})\left[\frac{\alpha_{s}({\rm 1 GeV})}{\alpha_{s}(\mu)}\right]^{\frac{12}{33-2n_f}}\, , \nonumber\\
 \langle\bar{q}g_s \sigma Gq \rangle(\mu)&=&\langle\bar{q}g_s \sigma Gq \rangle({\rm 1 GeV})\left[\frac{\alpha_{s}({\rm 1 GeV})}{\alpha_{s}(\mu)}\right]^{\frac{2}{33-2n_f}}\, ,\nonumber\\
  \langle\bar{s}g_s \sigma Gs \rangle(\mu)&=&\langle\bar{s}g_s \sigma Gs \rangle({\rm 1 GeV})\left[\frac{\alpha_{s}({\rm 1 GeV})}{\alpha_{s}(\mu)}\right]^{\frac{2}{33-2n_f}}\, ,\nonumber\\
m_c(\mu)&=&m_c(m_c)\left[\frac{\alpha_{s}(\mu)}{\alpha_{s}(m_c)}\right]^{\frac{12}{33-2n_f}} \, ,\nonumber\\
m_s(\mu)&=&m_s({\rm 2GeV} )\left[\frac{\alpha_{s}(\mu)}{\alpha_{s}({\rm 2GeV})}\right]^{\frac{12}{33-2n_f}}\, ,\nonumber\\
\alpha_s(\mu)&=&\frac{1}{b_0t}\left[1-\frac{b_1}{b_0^2}\frac{\log t}{t} +\frac{b_1^2(\log^2{t}-\log{t}-1)+b_0b_2}{b_0^4t^2}\right]\, ,
\end{eqnarray}
  where $t=\log \frac{\mu^2}{\Lambda^2}$, $b_0=\frac{33-2n_f}{12\pi}$, $b_1=\frac{153-19n_f}{24\pi^2}$, $b_2=\frac{2857-\frac{5033}{9}n_f+\frac{325}{27}n_f^2}{128\pi^3}$,  $\Lambda_{QCD}=210\,\rm{MeV}$, $292\,\rm{MeV}$  and  $332\,\rm{MeV}$ for the flavors  $n_f=5$, $4$ and $3$, respectively  \cite{PDG}.
As we investigate   the triply-charm  tetraquark states,  we choose the  flavor numbers $n_f=4$, then evolve all the input parameters to a pertinent  energy scale $\mu$ to compute  the tetraquark  masses.

The hidden-charm/bottom tetraquark states could be understood  by  a double-well potential model,  the heavy  quark $Q$ (antiquark $\overline{Q}$) serves as an  almost static well potential and attracts  the light quark $q$ (antiquark $\bar{q}$),  a diquark (antidiquark) in the $\bar{3}_c$ ($3_c$) comes into being. The cloud of the light quark $q$ dresses the heavy quark $Q$ to contribute an effective heavy quark mass ${\mathbb{M}}_Q$, then we obtain   the virtuality $V=\sqrt{M^2_{X/Y/Z}-(2{\mathbb{M}}_Q)^2}$, and set
 $ \mu^2=V^2={\mathcal{O}}(T^2)$
to reach the energy scale formula \cite{WangZG-review-2026,Wang-tetra-formula},
\begin{eqnarray}
\mu &=&\sqrt{M^2_{X/Y/Z}-(2{\mathbb{M}}_Q)^2}\, .
\end{eqnarray}
If we take it for granted  that the masses of the triply-charm tetraquark states meet with the energy scale formula, and choose the updated effective $c$-quark mass ${\mathbb{M}}_c=1.82\,\rm{GeV}$ to estimate the (lowest) ground state mass \cite{WangZG-eff-Mc},
\begin{eqnarray}
M_{Z}&=&\sqrt{(3{\mathbb{M}}_c)^2+({\rm 1 \,GeV})^2}\, ,\nonumber\\
 &=&5.55\,\rm{GeV}\, ,
\end{eqnarray}
which might overestimate the lowest mass. In Ref.\cite{Tetra-ccc-QCDSR-ZhangWS-NPA-2025}, Zhang and Tang studied the $\bar{3}_c3_c$-type $[cc]_{A}[\bar{c}\bar{q}]_{A}$, $[cc]_{A}[\bar{c}\bar{s}]_A$    and   $6_c\bar{6}_c$-type $[cc]_{S/P/V}[\bar{c}\bar{q}]_{S/P/V}$, $[cc]_{S/P/V}[\bar{c}\bar{s}]_{S/P/V}$ tetraquark states with the spin-parity  $J^P=0^+$, and adopted the energy scale formula to select the suitable  energy scales, and obtained tetraquark  masses larger than $5.55\,\rm{GeV}$.
In the present case, the $c$-quark attracts another $c$-quark to form an $A$ or $T$ diquark in the color anti-triplet, the two $c$ quarks dress each other and provide  an effective  doubly-charm diquark mass ${\mathbb{M}}_{cc}$, which could be estimated as half of the centroid of the ground state fully-charm tetraquark states \cite{WangZG-NPB-X6660},
\begin{eqnarray}
2{\mathbb{M}}_{cc}&=&\frac{M_{0^{++}}+3M_{1^{+-}}+5M_{2^{++}}}{9}\, ,\nonumber\\
&=&6.25\,\rm{GeV}\, .
\end{eqnarray}
Then we estimate the (lowest) ground state mass,
\begin{eqnarray}
M_{Z}&=&\sqrt{({\mathbb{M}}_c+{\mathbb{M}}_{cc})^2+({\rm 1 \,GeV})^2}\, ,\nonumber\\
 &=&5.05\,\rm{GeV}\, .
\end{eqnarray}
The lowest ground state mass might lie in the range $5.05\sim 5.55\,\rm{GeV}$.
In the present work, we abandon the constraint $\mu=\sqrt{M^2_{Z}-(3{\mathbb{M}}_c)^2}$ and take such an estimated ground state mass as a guide and search for the best energy scale of the QCD spectral densities, and finally we obtain the value $\mu=1.2\,\rm{GeV}$.

We cannot determine the pertinent continuum threshold parameters by the QCD sum rules themselves completely, and often consult the experimental data as a guide.
We  tentatively assign the  $Z^\pm_c(4430)$ as the 2S state of the $Z^\pm_c(3900)$ considering  the
analogous decays,
$Z_c^\pm(3900)\to J/\psi\pi^\pm$, $Z_c^\pm(4430)\to\psi^\prime\pi^\pm$
and analogous mass gaps  $M_{Z_c(4430)}-M_{Z_c(3900)}=591\,\rm{MeV}$ and $M_{\psi^\prime}-M_{J/\psi}=589\,\rm{MeV}$ from the Particle Data Group \cite{PDG,Maiani-Z4430-1405,Nielsen-1401,WangZG-Z4430-CTP,WangZG-Tetra-HC-PRD-2020}.
Furthermore,  we tentatively assign the $X(3915)$ and $X(4500)$ as the 1S and 2S hidden-charm-hidden-strange  tetraquark states with the spin-parity-change-conjugation $J^{PC}=0^{++}$ \cite{X4140-tetraquark-Lebed,X3915-X4500-EPJC-WZG,WangZG-Tetra-HC-HS-NPB-2024},  assign the $Z_c(4020)$ and $Z_c(4600)$ as the 1S and 2S hidden-charm tetraquark states with the spin-parity-change-conjugation $J^{PC}=1^{+-}$, respectively  \cite{ChenHX-Z4600-A,WangZG-axial-Z4600},  assign the $X(4140)$ and $X(4685)$ as the 1S and 2S hidden-charm-hidden-strange  tetraquark states with the spin-parity-change-conjugation $J^{PC}=1^{++}$, respectively \cite{WangZG-Tetra-HC-HS-NPB-2024,WZG-Di-X4140-EPJC,WZG-X4140-X4685}, and assign the $Y(4260)$ and $Y(4750)$ as the 1P and 2P hidden-charm  tetraquark states with the spin-parity-change-conjugation $J^{PC}=1^{--}$, respectively \cite{WangZG-Tetra-Y4260-Y4750-CPC-2024}. If such assignments make sense,  the energy (mass)  gaps between the 1S and 2S tetraquark states  are about $0.57\sim 0.59 \,\rm{GeV}$, we would like to choose the continuum threshold parameters as $\sqrt{s_0}=M_Z+0.55\sim0.59\pm 0.10\,\rm{GeV}$ for the triply-charm tetraquark states  to take account of the ground state contributions as large as possible.

The pole dominance on the hadron  side and convergence of the operator product expansion on the QCD side are two fundamental    criteria, we should fulfill  the two fundamental  criteria to reach  robust  QCD sum rules.
We define the pole contributions (PC) as usual,
\begin{eqnarray}
{\rm{PC}}&=&\frac{\int_{9m_{c}^{2}}^{s_{0}}
ds\,\rho_{QCD}\left(s\right)\exp\left(-\frac{s}{T^{2}}\right)} {\int_{9m_{c}^{2}}^{\infty}ds\,\rho_{QCD}\left(s\right)\exp\left(-\frac{s}{T^{2}}\right)}\, ,
\end{eqnarray}
 and define the contributions of the vacuum condensates  of dimension $n$ also as usual,
\begin{eqnarray}
D(n)&=&\frac{\int_{9m_{c}^{2}}^{s_{0}}ds\,\rho_{QCD,n}(s)\exp\left(-\frac{s}{T^{2}}\right)}
{\int_{9m_{c}^{2}}^{s_{0}}ds\,\rho_{QCD}\left(s\right)\exp\left(-\frac{s}{T^{2}}\right)}\, .
\end{eqnarray}

After numerous trial and error, we obtain the satisfactory  Borel windows, continuum threshold parameters,   pole contributions, and $D(7)$, which are shown plainly  in Table \ref{BorelP}.
On the hadron side, the pole contributions are about $(50-75)\%$ for all the twenty-four channels, the pole dominance criterion   is  satisfied very well.
On the QCD side, the contributions $|D(7)|\leq 1 \%$ or $\ll 1\%$, except that $|D(7)|<  2\%$ for the $[cc]_{\tilde{A}}[\overline{cu}]_{\tilde{A}}$ state with  the $J^P=0^{+}$,
$|D(7)|< 4\%$ for the
$[cc]_{\tilde{V}}[\overline{cu}]_{S}$ state with the $J^P=1^{-}$,
$|D(7)| \leq 2\%$ for the
$[cc]_{\tilde{V}}[\overline{cs}]_{S}$ state with the $J^P=1^{-}$, all in all,  the  operator product  expansion converges  very well.

We take  account of all the uncertainties of the input parameters and obtain the masses and pole residues of the scalar, axialvector, vector and tensor triply-charm  tetraquark states, which are shown plainly in Table \ref{mass-Table}. From the Table, we could obtain the conclusion tentatively that the energy gaps between the $[cc][\bar{c}\bar{q}]$ and $[cc][\bar{c}\bar{s}]$ tetraquark states, in other words, the light-flavor $SU(3)$ breaking effects,  are about $0.09\,\rm{GeV}$, and of the same magnitude of the $\overline{MS}$ mass $m_s(\rm 2\,GeV)$.

\begin{figure}
 \centering
 \includegraphics[totalheight=6cm,width=7cm]{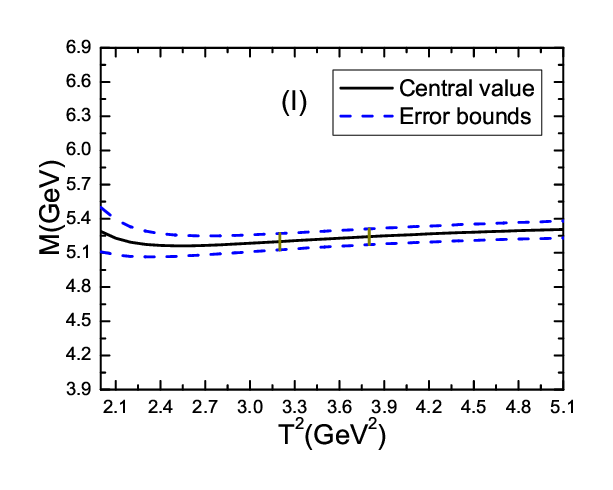}
 \includegraphics[totalheight=6cm,width=7cm]{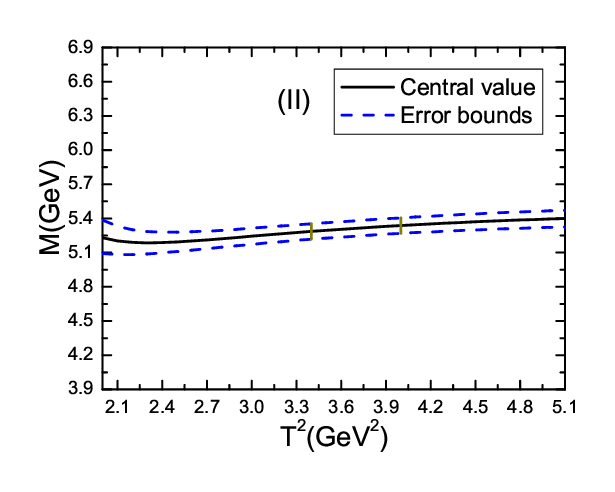}
 \caption{ The masses of the  $[cc]_{A}[\overline{cu}]_{S}$ (I) and $[cc]_{A}[\overline{cs}]_{S}$ (II)  tetraquark states   with variations  of the Borel parameters $T^2$.  }\label{mass-1-fig}
\end{figure}

 In  Fig.\ref{mass-1-fig},  as an example, we plot the predicted  masses of the  $[cc]_{A}[\overline{cu}]_{S}$  and $[cc]_{A}[\overline{cs}]_{S}$   tetraquark states with the spin-parity $J^P=1^+$ via variations of the Borel parameters at much larger ranges than the Borel widows, which are indicted by the two short vertical lines. From the figure, we can see clearly that there appear platforms in the Borel windows indeed.

In Ref.\cite{Tetra-ccc-QCDSR-ChenW-PRD-2017}, Jiang, Chen and Zhu studied the $\bar{3}_c3_c$ type $[cc]_{A}[\bar{c}\bar{q}]_{A}$ and $[cc]_{A}[\bar{c}\bar{q}]_{S}$ tetraquark states with the spin-parity $J^P=0^+$ and $1^+$, respectively,   and the  $6_c\bar{6}_c$ type $[cc]_{S}[\bar{c}\bar{q}]_{S}$ and $[cc]_{S}[\bar{c}\bar{q}]_{A}$ tetraquark states with the spin-parity $J^P=0^+$ and $1^+$, respectively,  and   obtained the degenerated  masses  $5.1\pm 0.2\,\rm{GeV}$, which are smaller than the present calculations, i.e. we obtain the mass $5.22\pm0.09\, \rm{GeV}$ for the $\bar{3}_c3_c$ type $[cc]_{A}[\overline{cu}]_{S}$ tetraquark state   with the spin-parity $J^P=1^{+}$, see Table \ref{mass-Table}.

In Ref.\cite{Tetra-ccc-QCDSR-LiuYL-2506}, Xu et al studied the $\bar{3}_c3_c$ type $[cc]_{A}[\bar{c}\bar{s}]_{A}$ and $[cc]_{\tilde{A}}[\bar{c}\bar{s}]_{\tilde{A}}$ tetraquark states with the spin-parity  $J^P=0^+$ via the QCD sum rules, and obtained the masses $5.0774^{+0.0708}_{-0.0641}~\mbox{GeV}$ and $5.0679^{+0.0839}_{-0.0721}~\mbox{GeV}$ respectively, which are also smaller than the present calculations.

We expect to compare the predicted triply-charm tetraquark states   to the experimental data  in the future to make contributions in searching for exotic states. Although the mass is the  most important benchmark in identifying  a hadron, we cannot identify a hadron unambiguously according to the mass alone, we  have to investigate  its productions and decays in a systematic way.  We can take the pole residues  as  input parameters to investigate   the two-body
 strong decays with the three-point QCD sum rules or light-cone QCD sum rules,
\begin{eqnarray}
Z_{ccc\bar{q}}(0^+)&\to& J/\psi D^*\, ,\, \eta_cD\, ,  \nonumber\\
Z_{ccc\bar{s}}(0^+)&\to& J/\psi D_s^*\, ,\, \eta_cD_s\, ,  \nonumber\\
Z_{ccc\bar{q}}(1^+)&\to& J/\psi D\, ,\,J/\psi D^*\, , \, \eta_cD^*\, ,  \nonumber\\
Z_{ccc\bar{s}}(1^+)&\to& J/\psi D_s\, ,\,J/\psi D_s^*\, , \, \eta_cD_s^*\, ,  \nonumber\\
Z_{ccc\bar{q}}(2^+)&\to& J/\psi D\, ,\,J/\psi D^*\, , \, \eta_cD^*\, , \, \eta_cD\, ,   \nonumber\\
Z_{ccc\bar{s}}(2^+)&\to& J/\psi D_s\, ,\,J/\psi D_s^*\, , \, \eta_cD_s^*\, , \, \eta_cD_s\, ,
\end{eqnarray}
to obtain additional support. 
It is not difficult to  construct the final states experimentally according to the decays $J/\psi \to \mu^+\mu^-$, $D^+ \to K^-\pi^+\pi^+$, $D^0 \to K^-\pi^+$, $K^-\pi^+\pi^-\pi^+$, $D^+_s \to K^+K^-\pi^+$, $\phi \pi^+$, $D^{*+}_s \to D^+_s \gamma$, $D^{*+} \to D^0\pi^+$, etc.
The LHCb collaboration have observed (evidences of) the decays $B^+_c \rightarrow J/\psi D_s^+$, $ J/\psi D_s^{*+}$, $ J/\psi D^0 K^+$, $ J/\psi D^{*0} K^+$, $ J/\psi D^{*+} K^{*0}$, $ J/\psi D^+ K^{*0}$, $J/\psi D^{\ast+}$  \cite{LHCb-JpsiD-PRD-2013,LHCb-JpsiD-PRD-2017,LHCb-JpsiDv-2026}.

\begin{table}
\begin{center}
\begin{tabular}{|c|c|c|c|c|c|c|c|c|}\hline\hline
 $Z_{cc\bar{c}}$              &$J^{P}$  &$T^2 (\rm{GeV}^2)$ &$\sqrt{s_0}(\rm GeV) $         &pole         &$|D(7)|$ \\ \hline

$[cc]_{A}[\overline{cu}]_{A}$      &$0^{+}$ &$3.2-3.8$       &$5.84\pm0.10$                           &$(49-74)\%$  &$\le 1\%$    \\

$[cc]_{\tilde{A}}[\overline{cu}]_{\tilde{A}}$      &$0^{+}$ &$3.1-3.7$       &$5.87\pm0.10$                           &$(50-75)\%$  &$< 2\%$    \\ \hline

$[cc]_{A}[\overline{cs}]_{A}$      &$0^{+}$ &$3.3-3.9$       &$5.95\pm0.10$                           &$(51-76)\%$  &$< 1\%$    \\

$[cc]_{\tilde{A}}[\overline{cs}]_{\tilde{A}}$      &$0^{+}$ &$3.3-3.9$       &$5.98\pm0.10$              &$(49-74)\%$  &$\leq 1\%$    \\ \hline

$[cc]_{A}[\overline{cu}]_{S}$      &$1^{+}$ &$3.2-3.8$       &$5.79\pm0.10$                           &$(51-75)\%$  &$\ll 1\%$    \\

$[cc]_{\tilde{A}}[\overline{cu}]_{A}$      &$1^{+}$ &$3.3-3.9$  &$5.89\pm0.10$ &$(49-73)\%$  &$< 1\%$    \\

$[cc]_{A}[\overline{cu}]_{A}$      &$1^{+}$ &$3.4-4.0$       &$5.89\pm0.10$                           &$(50-74)\%$  &$\ll 1\%$    \\

$[cc]_{\tilde{A}}[\overline{cu}]_{\tilde{A}}$      &$1^{+}$ &$3.3-3.9$       &$5.92\pm0.10$                           &$(49-74)\%$  &$\sim 1\%$    \\

$[cc]_{\tilde{A}}[\overline{cu}]_{S}$      &$1^{+}$ &$3.3-3.9$       &$5.79\pm0.10$                           &$(49-73)\%$  &$\ll 1\%$    \\ \hline

$[cc]_{A}[\overline{cs}]_{S}$      &$1^{+}$ &$3.4-4.0$       &$5.90\pm0.10$                           &$(50-74)\%$  &$\ll 1\%$    \\

$[cc]_{\tilde{A}}[\overline{cs}]_{A}$      &$1^{+}$ &$3.5-4.1$       &$6.00\pm0.10$                           &$(49-73)\%$  &$< 1\%$    \\

$[cc]_{A}[\overline{cs}]_{A}$      &$1^{+}$ &$3.6-4.2$       &$6.00\pm0.10$                           &$(50-73)\%$  &$\ll 1\%$    \\

$[cc]_{\tilde{A}}[\overline{cs}]_{\tilde{A}}$      &$1^{+}$ &$3.5-4.1$       &$6.03\pm0.10$                           &$(50-73)\%$  &$\leq 1\%$    \\

$[cc]_{\tilde{A}}[\overline{cs}]_{S}$      &$1^{+}$ &$3.4-4.0$       &$5.90\pm0.10$                           &$(51-74)\%$  &$\ll 1\%$    \\
\hline

$[cc]_{A}[\overline{cu}]_{A}$      &$1^{-}$ &$4.1-4.9$       &$6.61\pm0.10$                           &$(50-74)\%$  &$\le 1\%$    \\

$[cc]_{\tilde{A}}[\overline{cu}]_{\tilde{A}}$      &$1^{-}$ &$4.1-4.9$       &$6.57\pm0.10$                           &$(50-74)\%$  &$< 1\%$    \\

$[cc]_{\tilde{V}}[\overline{cu}]_{S}$      &$1^{-}$ &$3.3-3.9$       &$6.09\pm0.10$                           &$(51-75)\%$  &$< 4\%$    \\ \hline

$[cc]_{A}[\overline{cs}]_{A}$      &$1^{-}$ &$4.3-5.1$       &$6.72\pm0.10$                           &$(51-74)\%$  &$< 1\%$    \\

$[cc]_{\tilde{A}}[\overline{cs}]_{\tilde{A}}$      &$1^{-}$ &$4.3-5.1$       &$6.68\pm0.10$                           &$(50-73)\%$  &$< 1\%$    \\

$[cc]_{\tilde{V}}[\overline{cs}]_{S}$      &$1^{-}$ &$3.5-4.1$       &$6.20\pm0.10$                           &$(50-74)\%$  &$\leq 2\%$    \\
\hline

$[cc]_{A}[\overline{cu}]_{A}$      &$2^{+}$ &$3.5-4.1$       &$5.97\pm0.10$                           &$(51-74)\%$  &$\ll 1\%$    \\

$[cc]_{\tilde{A}}[\overline{cu}]_{\tilde{A}}$      &$2^{+}$ &$3.3-3.9$       &$5.99\pm0.10$                           &$(51-75)\%$  &$< 1\%$    \\ \hline

$[cc]_{A}[\overline{cs}]_{A}$      &$2^{+}$ &$3.7-4.3$       &$6.08\pm0.10$                           &$(51-73)\%$  &$\ll 1\%$    \\

$[cc]_{\tilde{A}}[\overline{cs}]_{\tilde{A}}$      &$2^{+}$ &$3.6-4.2$       &$6.10\pm0.10$                           &$(50-73)\%$  &$< 1\%$    \\

\hline\hline
\end{tabular}
\end{center}
\caption{ The Borel parameters, continuum threshold parameters,  pole contributions, and  $D(7)$  for the ground state triply-charm tetraquark states. }\label{BorelP}
\end{table}

\begin{table}
\begin{center}
\begin{tabular}{|c|c|c|c|c|c|c|c|c|}\hline\hline
$Z_{cc\bar{c}}$      &$J^{P}$  &$M_Z (\rm{GeV})$   &$\lambda_Z (\rm{GeV}^5) $             \\ \hline

$[cc]_{A}[\overline{cu}]_{A}$    &$0^{+}$  &$5.29\pm0.09$  &$(1.56\pm0.26)\times 10^{-1}$         \\

$[cc]_{\tilde{A}}[\overline{cu}]_{\tilde{A}}$    &$0^{+}$  &$5.31\pm0.09$  &$(3.25\pm0.59)\times 10^{-1}$         \\ \hline

$[cc]_{A}[\overline{cs}]_{A}$    &$0^{+}$  &$5.37\pm0.10$  &$(1.81\pm0.32)\times 10^{-1}$         \\

$[cc]_{\tilde{A}}[\overline{cs}]_{\tilde{A}}$    &$0^{+}$  &$5.40\pm0.10$  &$(3.85\pm0.72)\times 10^{-1}$         \\ \hline

$[cc]_{A}[\overline{cu}]_{S}$    &$1^{+}$  &$5.22\pm0.09$  &$(8.01\pm1.34)\times 10^{-2}$         \\

$[cc]_{\tilde{A}}[\overline{cu}]_{A}$    &$1^{+}$  &$5.33\pm0.09$  &$(1.32\pm 0.23)\times 10^{-1}$         \\

$[cc]_{A}[\overline{cu}]_{A}$    &$1^{+}$  &$5.34\pm0.09$  &$(9.05\pm1.41)\times 10^{-2}$         \\

$[cc]_{\tilde{A}}[\overline{cu}]_{\tilde{A}}$    &$1^{+}$  &$5.36\pm0.09$  &$(9.54\pm1.59)\times 10^{-2}$         \\

$[cc]_{\tilde{A}}[\overline{cu}]_{S}$    &$1^{+}$  &$5.22\pm0.09$  &$(7.79\pm1.30)\times 10^{-2}$         \\ \hline

$[cc]_{A}[\overline{cs}]_{S}$    &$1^{+}$  &$5.31\pm0.10$  &$(9.45\pm1.64)\times 10^{-2}$         \\

$[cc]_{\tilde{A}}[\overline{cs}]_{A}$    &$1^{+}$  &$5.42\pm0.10$  &$(1.56\pm 0.28)\times 10^{-1}$         \\

$[cc]_{A}[\overline{cs}]_{A}$    &$1^{+}$  &$5.42\pm0.09$  &$(1.07\pm0.17)\times 10^{-1}$         \\

$[cc]_{\tilde{A}}[\overline{cs}]_{\tilde{A}}$    &$1^{+}$  &$5.45\pm0.10$  &$(1.12\pm0.19)\times 10^{-1}$         \\

$[cc]_{\tilde{A}}[\overline{cs}]_{S}$    &$1^{+}$  &$5.31\pm0.10$  &$(9.11\pm1.56)\times 10^{-2}$         \\ \hline

$[cc]_{A}[\overline{cu}]_{A}$    &$1^{-}$  &$6.04\pm0.11$  &$(1.58\pm0.26)\times 10^{-1}$         \\

$[cc]_{\tilde{A}}[\overline{cu}]_{\tilde{A}}$    &$1^{-}$  &$5.99\pm0.11$  &$(2.21\pm0.37)\times 10^{-1}$         \\

$[cc]_{\tilde{V}}[\overline{cu}]_{S}$    &$1^{-}$  &$5.53\pm0.10$  &$(5.45\pm0.93)\times 10^{-2}$         \\ \hline

$[cc]_{A}[\overline{cs}]_{A}$    &$1^{-}$  &$6.14\pm0.11$  &$(1.79\pm0.27)\times 10^{-1}$         \\

$[cc]_{\tilde{A}}[\overline{cs}]_{\tilde{A}}$    &$1^{-}$  &$6.09\pm0.10$  &$(2.52\pm0.38)\times 10^{-1}$         \\

$[cc]_{\tilde{V}}[\overline{cs}]_{S}$    &$1^{-}$  &$5.62\pm0.10$  &$(6.43\pm1.15)\times 10^{-2}$         \\ \hline

$[cc]_{A}[\overline{cu}]_{A}$    &$2^{+}$  &$5.41\pm0.09$  &$(1.44\pm0.21)\times 10^{-1}$         \\

$[cc]_{\tilde{A}}[\overline{cu}]_{\tilde{A}}$    &$2^{+}$  &$5.43\pm0.09$  &$(1.50\pm0.24)\times 10^{-1}$         \\ \hline

$[cc]_{A}[\overline{cs}]_{A}$    &$2^{+}$  &$5.49\pm0.09$  &$(1.68\pm0.26)\times 10^{-1}$         \\

$[cc]_{\tilde{A}}[\overline{cs}]_{\tilde{A}}$    &$2^{+}$  &$5.52\pm0.09$  &$(1.77\pm0.29)\times 10^{-1}$         \\
\hline\hline
\end{tabular}
\end{center}
\caption{ The masses and pole residues of the ground state triply-charm tetraquark states. }\label{mass-Table}
\end{table}

\section{Conclusion}
In this work,  we take the   axialvector and tensor doubly-charm diquarks and scalar, axialvector, and tensor singly-charm anti-diquarks as the fundamental  constituents, and construct
  the scalar, axialvector and tensor triply-charm four-quark currents. Then we investigate  the  mass spectrum of the ground state triply-charm tetraquark states  with the spin-parity  $J^P=0^+$, $1^+$ and $2^+$ via the QCD sum rules comprehensively. Routinely, we accomplish  the operator product expansion up to the vacuum condensates of dimension $7$ according to our usual counting roles for consistence to reach  the QCD spectral densities, then we select the optimal energy scales, Borel parameters and continuum threshold parameters to meet with the two fundamental  criteria via numerous trial and error. At last, we obtain the masses and pole residues of the eighteen triply-charm tetraquark states with the spin-parity $J^P=0^+$, $1^+$ and $2^+$. As a byproduct, we also obtain the masses and pole residues of four triply-charm tetraquark states with the spin-parity $J^P=1^-$.
We would like to compare the predicted tetraquark states to the experimental data in the future and select the possible identifications of the exotic states, and take the pole residues  as  input parameters to investigate  the two-body  strong decays with the three-point QCD sum rules or light-cone QCD sum rules to reach additional support.

\section*{Acknowledgements}
This  work is supported by National Natural Science Foundation, Grant Number  12575083.

\end{document}